\documentclass[aps,prl,reprint,superscriptaddress]{revtex4-1}
\usepackage{cases}
\usepackage{xcolor}
\usepackage{braket}
\usepackage{graphicx}
\usepackage{longtable}

\begin{document}


\title{Dynamic characterization of an alkali-ion battery as a source for laser-cooled atoms}


\author{J. P. McGilligan}
\affiliation{University of Colorado, Department of Physics, Boulder, Colorado, 80309, USA}
\affiliation{National Institute of Standards and Technology, Boulder Colorado, 80305, USA}
\email{james.mcgilligan@nist.gov}
\author{K. R. Moore}
\affiliation{National Institute of Standards and Technology, Boulder Colorado, 80305, USA}
\author{S. Kang}
\affiliation{University of Colorado, Department of Physics, Boulder, Colorado, 80309, USA}
\affiliation{National Institute of Standards and Technology, Boulder Colorado, 80305, USA}
\author{R. Mott}
\author{A. Mis}
\author{C. Roper}
\affiliation{HRL Laboratories, LLC; Malibu, California, 90265, USA}
\author{E. A. Donley}
\author{J. Kitching}
\affiliation{National Institute of Standards and Technology, Boulder Colorado, 80305, USA}



\date{\today}

\begin{abstract}
We investigate a solid-state, reversible, alkali-ion battery (AIB) capable of regulating the density of alkali atoms in a vacuum system used for the production of laser-cooled atoms. The cold-atom sample can be used with in-vacuum chronoamperometry as a diagnostic for the voltage-controlled electrochemical reaction that sources or sinks alkali atoms into the vapor. In a combined reaction-diffusion-limited regime, we show that the number of laser-cooled atoms in a magneto-optical trap can be increased both by initially loading the AIB from the vapor for longer, and by using higher voltages across the AIB when atoms are subsequently sourced back into the vapor. The time constants associated with the change in atom number in response to a change in AIB voltage are in the range of 0.5~s~-~40~s. The AIB alkali reservoir is demonstrated to survive oxidization during atmospheric exposure, simplifying reservoir loading prior to vacuum implementation as a replacement for traditional resistively-heated dispensers. The AIB capabilities may provide an improved atom number stability in next-generation atomic clocks and sensors, while also facilitating fast loading and increased interrogation times. 
\end{abstract}

\pacs{}

\maketitle

\section{Introduction}
Since the first demonstration of laser cooling, ensembles of laser-cooled neutral atoms have been implemented in a wide variety of precision instruments due to the long interrogation times that can be achieved without interaction with the environment. These properties are notably utilized in state-of-the-art metrological experiments, such as atom interferometers and atomic clocks \cite{ludlow, Katori, kasevich, tino, arnold}. The improved performance of these cold atom measurements compared to their thermal counterparts has led to significant efforts to miniaturize the laser cooling apparatus into a field-deployable device \cite{himsworth1,dellis1,mcgilligan1}.

Among the many challenges associated with moving cold atom instrumentation from the laboratory to the field is the stabilization of the alkali atom density. As the temperature of the environment changes, so too does the density of the alkali vapor from which the cold atom sample is loaded. The absence of an alkali density regulator will ultimately degrade the long-term magneto-optical trap (MOT) number stability, and can prevent a MOT from forming altogether for large enough deviations from room temperature. Techniques such as light-induced-atomic-desorption (LIAD) have demonstrated the ability to modulate and control the background alkali density \cite{cassettari, Karaulanov}, however such methods have not yet demonstrated fast desorption times, nor have they been miniaturized to facilitate portability and micro-fabrication \cite{songbai1}. 

Previously, we demonstrated stabilization of the alkali density in a vapor cell with a micro-fabricated, reversible, in-vacuum electrolytic device, and showed that alkali atoms sourced from such a device could be laser cooled and trapped in a MOT \cite{songbai1, songbai2}. In this paper we evaluate the performance of the AIB by using the MOT as an in-situ diagnostic tool to determine the characteristics of the electrochemical processes. We show that the Rb emission into the vapor depends on both the initial state-of-charge within the AIB and the applied voltage during sourcing. In doing so, we establish how the cold atom number and loading time depend on the operating parameters of the AIB and elucidate some of the underlying electrochemistry that determines the performance of the battery as a source for laser-cooled atomic ensembles.

\section{Experimental set-up}
\begin{figure}[t]
\centering
\includegraphics[width=1 \columnwidth]{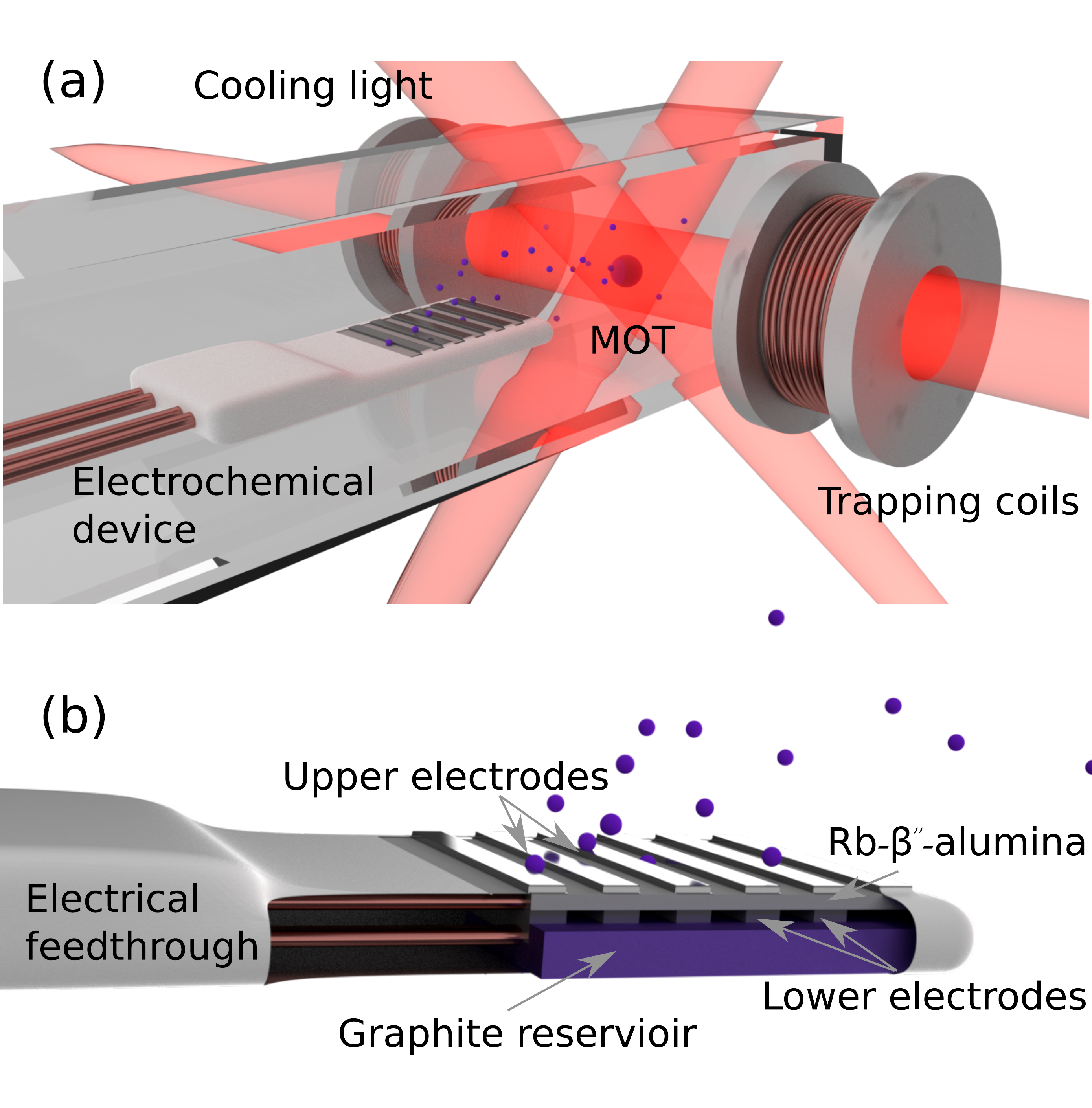}
\caption{(a) An illustration of the magneto-optical trap being sourced from the AIB. (b) A cut-through of the AIB is provided to reveal the internal composition.}
\label{setup}
\end{figure}

The experimental setup is illustrated in Fig.~\ref{setup} (a). We create a standard six-beam MOT close to the AIB, Fig.~\ref{setup} (b), with light from a single distributed-Bragg-reflector (DBR) laser, locked to address the $^{85}$Rb $F=3\rightarrow F'=4$ transition, while being frequency modulated at 2.9~GHz to generate $\approx$10$\%$ sidebands for re-pumping. The light is spatially filtered using an optical fiber and provides a total incident power of 35~mW, red-detuned by 2$\Gamma$, where $\Gamma$ is the natural linewidth of the transition, and a 1/e$^2$ diameter of 5~mm.
The vacuum system inside the 10~mm~$\times$~10~mm~$\times$~70~mm cell is maintained with a 2~L/s ion pump, and Rb can be introduced with standard heated alkali dispensers (Rb$_2$MoO$_4$/AlZr). The cell walls are coated in octadecyltrichlorosilane (OTS) to minimize alkali adsorption. 

The internal body of the AIB houses a layer of ion conducting Rb-$\beta$''-alumina, sandwiched between upper and lower electrodes. The upper electrode couples the Rb-$\beta$''-alumina to the outer environment via an array of electrode fingers that cover the upper surface. The lower electrode is housed within a graphite alkali reservoir, which stores neutral alkali atoms. The alkali vapor density is enhanced when voltage is applied across the electrodes \cite{gong, bernstein}. The bidirectional functionality to source/sink alkali atoms from the vapor is controlled by the voltage applied across the AIB and its temperature. Excluding the upper electrodes and exposed Rb-$\beta$''-alumina between top electrode features, the entire AIB surface is insulated and coated in a vacuum epoxy.

MOT image acquisition was carried out with fluorescence imaging on a CCD camera. The background Rb level was actively measured using absorption spectroscopy and subtracted from the fluorescence data to reduce the effect of the changing background Rb density levels.

\section{Results}

\begin{figure}[t]
\centering
\includegraphics[width=0.90 \columnwidth]{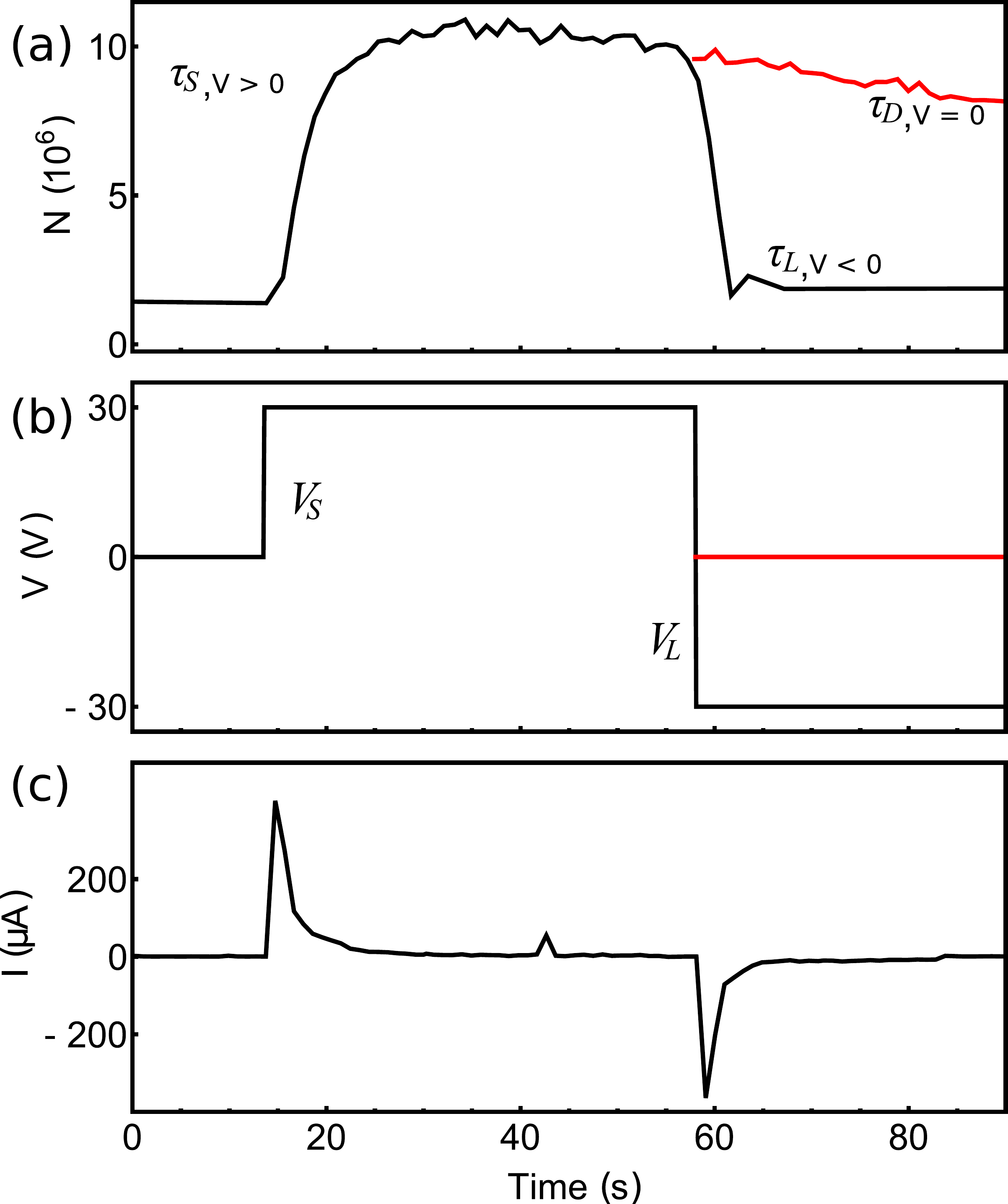}
\caption{(a) The number of laser-cooled atoms in the MOT, (b) the applied voltage, and (c) the corresponding current through the solid-state battery, as a function of time. At t~=~14~s, a positive voltage is applied to the battery causing current to flow and leading to an increase in the number of cold atoms with a time constant $\tau_{S}$. At t~=~58~s, the polarity of the applied voltage is reversed, leading to a reverse current and a decrease in the number of cold atoms with time constant $\tau_{L}$. If instead, the voltage is set to zero (red lines), the atom number decreases with the time constant $\tau_{D}$.}
\label{exampletau}
\end{figure}

The typical response of the measured MOT atom-number $N$ and AIB current $I$ to a change in the applied voltage polarity $V$ is illustrated in Fig.~\ref{exampletau}. The AIB is first loaded with alkali atoms through the application of a negative voltage for some duration of time before t~=~0~s in the presence of an alkali vapor in the cell provided by a standard heated alkali dispenser in the hours before the experiment. The MOT light is turned on and a small number of laser-cooled atoms appears. As a positive voltage step is applied at t~=~14~s, the number of cold atoms in the MOT is observed to increase and reaches steady-state with a time constant $\tau_{S}$ between 0.5~s and 40~s. We note that this is not the typical MOT loading curve observed in many cold atom experiments as a result of switching on a magnetic field gradient, described in Ref.~\cite{wieman}, but instead reflects changes in the alkali density induced by the AIB. This increase in the number of cold atoms is associated with a sharp increase in the device current. When the polarity of the applied voltage is reversed at t~=~58~s, the cold atom number decays to near the detection noise floor with time constant $\tau_{L}$ between 0.5~s and 10~s, again consistent with \cite{songbai3}. If instead the voltage is set to zero at t~=~58~s, a much slower decay of the atom number is observed, with decay time constant $\tau_{D}\approx$100~s.

The behaviour shown in Fig.~\ref{exampletau} can be explained as follows. When the positive voltage is applied, the neutral Rb at the interface of the lower electrode and graphite is ionized and conducted into the Rb-$\beta$''-alumina electrolyte as a result of the electric field created by the applied voltage. Rb$^+$ ions recombine with electrons at the interface between the upper electrodes and Rb-$\beta$"-alumina, producing a large initial current while neutral Rb metal accumulates on the surface. As the electro-chemical process depletes the Rb within the graphite, a Rb concentration gradient forms within the graphite reducing the Rb density near the electrode and limiting the rate of the electrochemical process by the diffusion rate of the remaining neutral Rb out of the graphitic reservoir \cite{brownson, molina}. The measured time-dependent current is observed in Fig.~\ref{exampletau} (c) to decay with a $1/\sqrt{t}$ dependence. In these measurements, only the flow of Faradaic current is analyzed \cite{molina, anson}. 

By integrating the current, the total charge $Q$ that has flowed between the upper and lower electrodes can be found. This charge represents the number of ions conducted through the Rb-$\beta$"-alumina, as permitted by electrochemical reactions at the lower and upper electrodes. Following the electrochemical process, the neutral Rb metal likely diffuses across the upper AIB surface to form a metallic film, and is evaporated into the vacuum.

It is interesting to note that the response of the MOT atom number at t~=~14~s is somewhat slower that the corresponding decay of the AIB current. We attribute this difference to the fact that in order for the alkali vapor density (and hence MOT atom number) to increase, the alkali metal likely diffuses on the AIB surface to form a film and subsequently evaporates into the vacuum chamber. The current flow through the AIB does not depend on these processes and if they take longer than the electrochemical processes, changes in the atomic vapor (and hence MOT atom number) would be expected to occur more slowly. 

At t~=~58~s, the polarity of the voltage is reversed and the device switches from being a source of alkali atoms to the vapor to being a sink from the vapor. Under these conditions, vapor-phase Rb atoms that strike the upper electrodes are ionized and flow through the Rb-$\beta$"-alumina to the reservoir due to the applied electric field. The Rb$^+$ ions undergo reduction at the lower electrode-graphite interface, and diffuse into the reservoir as neutral Rb until a source voltage is applied. If instead of sinking to the reservoir, the source voltage is simply turned off, the atom number decays at a much slower rate $\tau_{D}$. We speculate that this slower decay results from a slow decrease of the metallic Rb coverage on the AIB surface due to removal of Rb from the vapor (and hence AIB surface) by the ion pump. This reduces the rate at which alkali atoms vaporize from the surface and, combined with a constant pumping rate from the ion pump, results in a lower overall Rb density in the vapor.

\subsection{AIB loading time}

\begin{figure}[h!]
\centering
\includegraphics[width=0.92 \columnwidth]{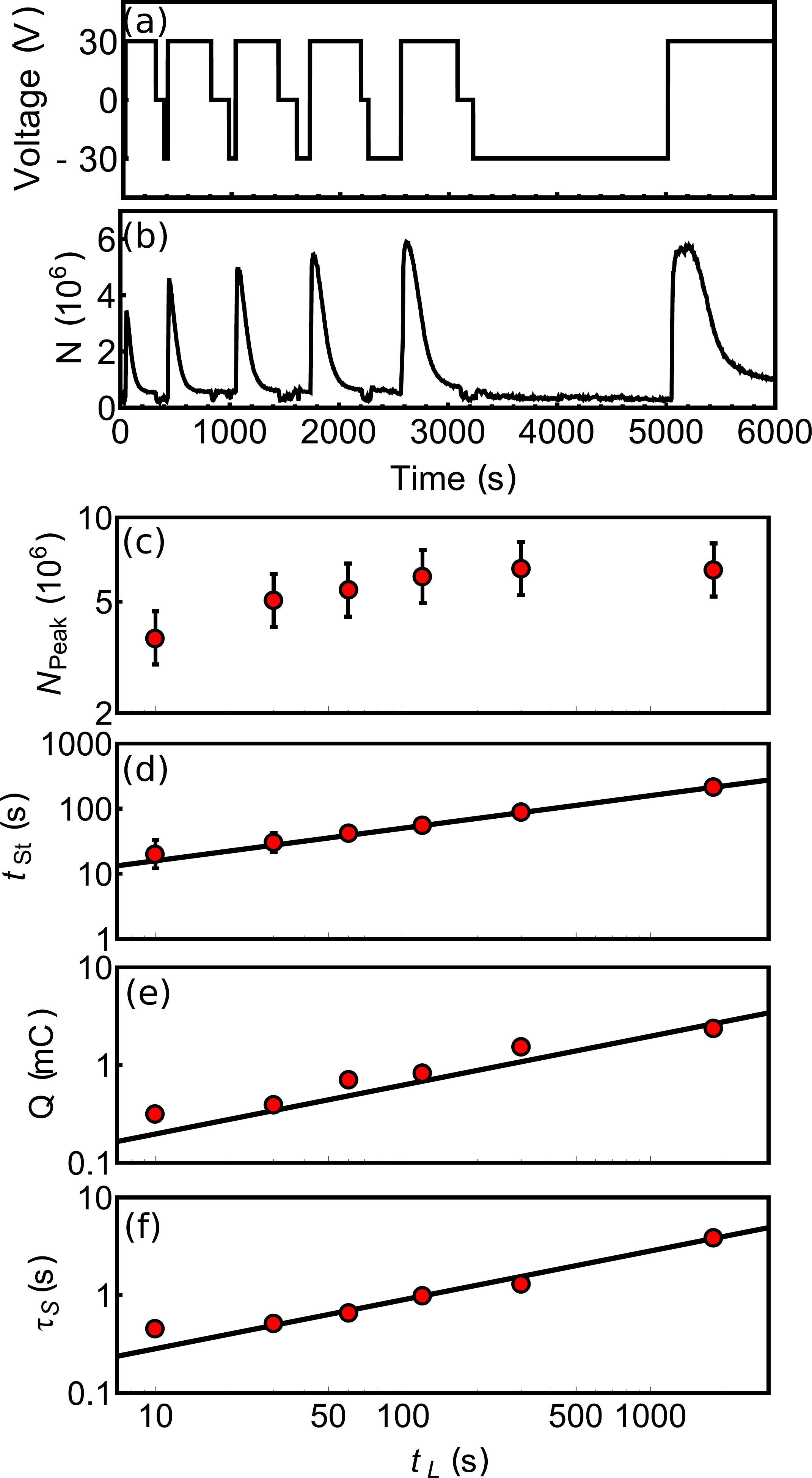}
\caption{Device performance as a function of the time during which the device loads atoms from the vapor. (a) The voltage applied across the device as a function of time showing five experimental cycles of loading at -30 V followed by sourcing at + 30 V. The time during which the device loads atoms at -30 V is increased with each subsequent cycle. (b) The number of cold atoms in the MOT increases for longer loading times. (c) The peak atom number, (d) the time the MOT number remains at steady-state, (e) charge transferred across the AIB (measured by integrating the current pulse), and (f) the time constant for the MOT atom number to reach steady-state after application of the sourcing voltage, $\tau_{S}$. Black lines represent power law fits to the data.}
\label{sinktime}
\end{figure}
The number of atoms in the MOT depends on the time over which atoms were initially loaded into the AIB, $t_{L}$. To ensure a repeatable initial state of the experiment, a positive voltage was applied across the AIB to deplete any previously stored Rb from the reservoir. Prior to the set of AIB loading time measurements, the alkali dispenser was heated to temporarily increase the Rb pressure in the vacuum chamber. For each loading time tested, Rb was loaded into the AIB by applying -30~V to the AIB electrodes for a variable time. Following loading, the voltage was briefly turned off, then the polarity was reversed and atoms were sourced from the AIB at 30~V while monitoring the fluorescence from the MOT. The applied voltage as a function of time and the corresponding number of atoms detected in the MOT are shown in Fig.~\ref{sinktime} (a) and (b). Fig.~\ref{sinktime} (c) shows that the peak atom number, $N_{Peak}$, measured for a constant source voltage, increases with $t_{L}$ between 10~s to 1000~s. We note that for the larger loading times there is indication that the reservoir concentration has saturated. The total time the atom number remains at steady-state $t_{St}$, defined by the time taken for the peak atom number to decrease by 25$\%$, was seen to be consistent with a $\sqrt{t_{L}}$ dependence, as shown in Fig.~\ref{sinktime} (d), where the black line is a power law fit. The total charge between the electrodes, determined by integrating the area under the current during the first 50~s of sourcing, was observed to increase as $\sqrt{t_{L}}$ over this same increase in loading time, as illustrated in Fig.~\ref{sinktime} (e). Fig.~\ref{sinktime} (f) illustrates that the time for the MOT number to reach steady-state, $\tau_{S}$, is also consistent with a $\sqrt{t_{L}}$ function of the loading time.

Longer loading times at the same loading voltage should result in a larger number of neutral Rb atoms entering the device, and hence a higher density of Rb within the reservoir. Such an increased Rb content produces a larger total charge during sourcing under the same sourcing voltage, leading to the formation of a larger number of Rb metal atoms on the surface and greater device surface coverage. This results in an increased evaporation rate and larger Rb vapor density and MOT number. The reason that $\tau_{S}$ increases with AIB loading time requires further investigation, but is possibly due to the atoms penetrating deeper within the reservoir. The measured increase of the sourced atom number and charge transfer indicates the peak atom number for a constant applied voltage is dictated by the bulk concentration within the reservoir. 

We found that Rb could be stored within the AIB reservoir during exposure to atmosphere. Rb was loaded into the AIB from a background alkali density before exposure to atmosphere for one hour. Following this exposure, the vacuum was reestablished below 10$^{-8}$ mbar in the chamber and when a sourcing voltage was applied to the AIB, the MOT could be loaded, demonstrating the ability to store pure metallic Rb in oxygen and water vapor containing atmosphere without complete Rb oxidation. 

\subsection{AIB sourcing voltage}
\begin{figure}[t]
\centering
\includegraphics[width=0.95 \columnwidth]{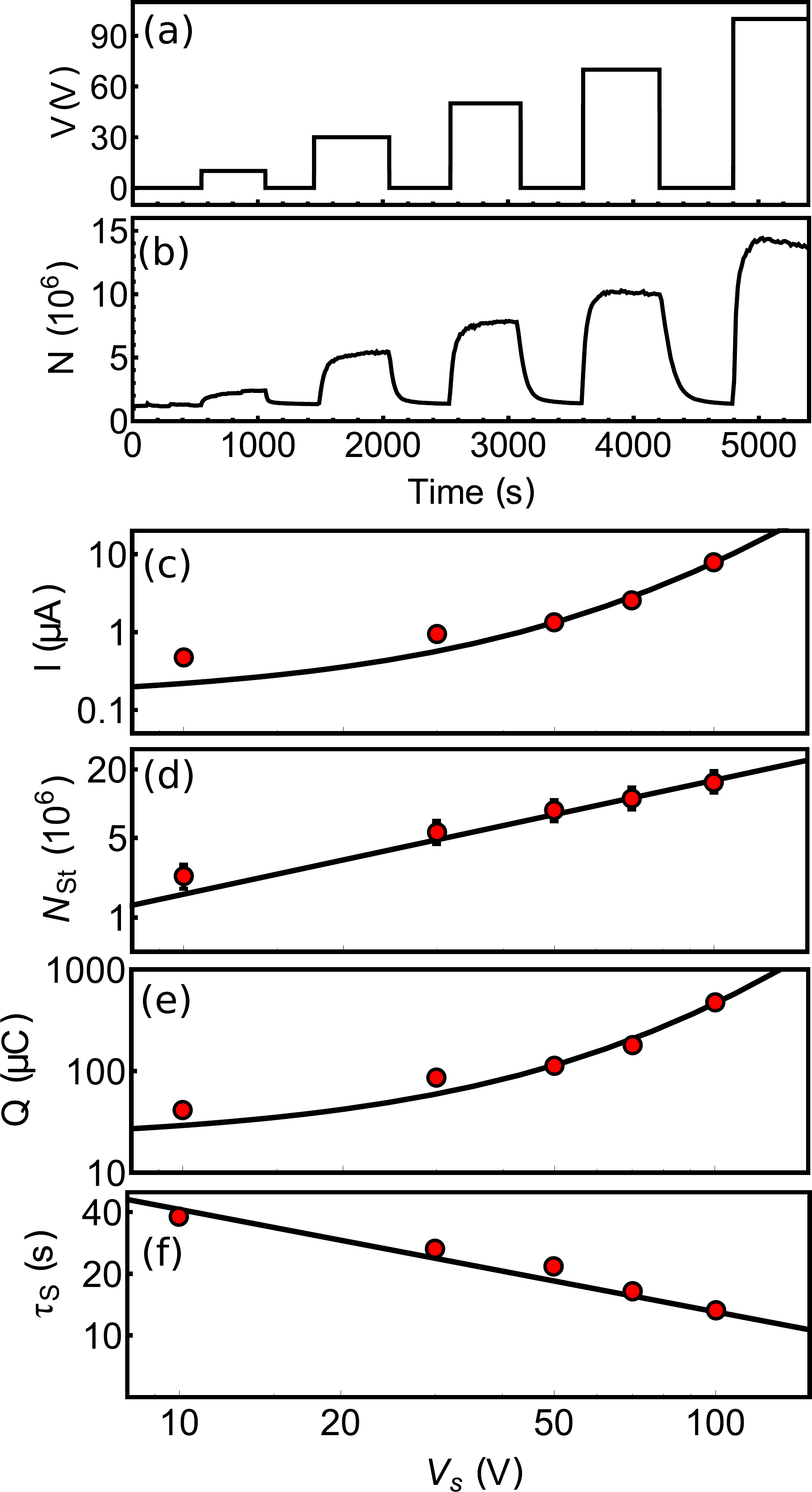}
\caption{Device performance as a function of the steady-state source voltage. (a): The raw time domain data of the applied voltage. (b): The raw time domain data of the atom number sourcing. (c): The steady-state current. (d): The steady-state MOT number. (e): The charge during the first 50~s sourcing. (f): The sourcing time constant, $\tau_{S}$. Black lines indicate the best fit to the data with the functional forms described in the text.}
\label{appliedI}
\end{figure}

The MOT atom number and loading time depended on the applied sourcing voltage for constant loading conditions. The voltage across the AIB was set to -30~V to load Rb from the vacuum for 30 minutes. Subsequently, positive voltage $V_s$ was applied as illustrated in Fig.~\ref{appliedI} (a), resulting in the formation of a MOT with a steady-state number $N_{St}$ and a time constant $\tau_{S}$, as shown in Fig.~\ref{appliedI} (b). The voltage was then set to 0~V and the background alkali density was allowed to fall back to its initial level. This procedure was repeated for five voltages.
For an electrochemical reaction contribution, the current is expected to have an exponential dependence on the applied voltage per the Butler-Volmer equation \cite{bard}. An exponential relation between the applied voltage and measured steady-state current was observed from this data set, illustrated in Fig.~\ref{appliedI} (c). The exponential relation indicates that when in the regime of an increasing voltage, the AIB is not entirely diffusion-limited and is instead in a regime where there is a significant contribution from the electrochemical reaction process. 

The steady-state atom numbers that were measured following sourcing are plotted in Fig.~\ref{appliedI} (d) as a function of the AIB applied voltage, where the black line represents the best power law fit. The results demonstrate a larger number of sourced Rb for higher applied voltages and higher currents. In the case of sourcing, we find the data is consistent with $N_{St}\propto V_{S}$. Again, the MOT number is demonstrated to be correlated with the Rb$^+$ charge into and out of the Rb-$\beta$"-alumina, illustrated in Fig.~\ref{appliedI} (e), where we see a an exponential increase in the charge over the range of voltages from 10~V to 100~V. 

The time constant $\tau_{S}$, shown in Fig.~\ref{appliedI} (f), decreases with $1/\sqrt{V_S}$, as highlighted by the black line power law fit. The faster time constant would be expected for a faster electrochemical recombination process generated by the higher potential at larger voltages. However, the long time constants observed here may be the result of a secondary process that increases $\tau_{S}$. The cause of this is not entirely known, but could possibly be caused by Rb diffusion on the top surface of the device prior to evaporation. In this scenario, the increased charge is due to the larger applied voltage during sourcing providing an increased potential to aid a faster electrochemical recombination rate at the surface. However, we note that although the total source time is less than the total load time, there may be an effect of reservoir depletion visible towards the end of the experiment, highlighted by the decreasing atom number from steady-state at t~=~5000~s in Fig.~\ref{appliedI} (b).

\subsection{AIB loading voltage}
\begin{figure}[t]
\centering
\includegraphics[width=0.95 \columnwidth]{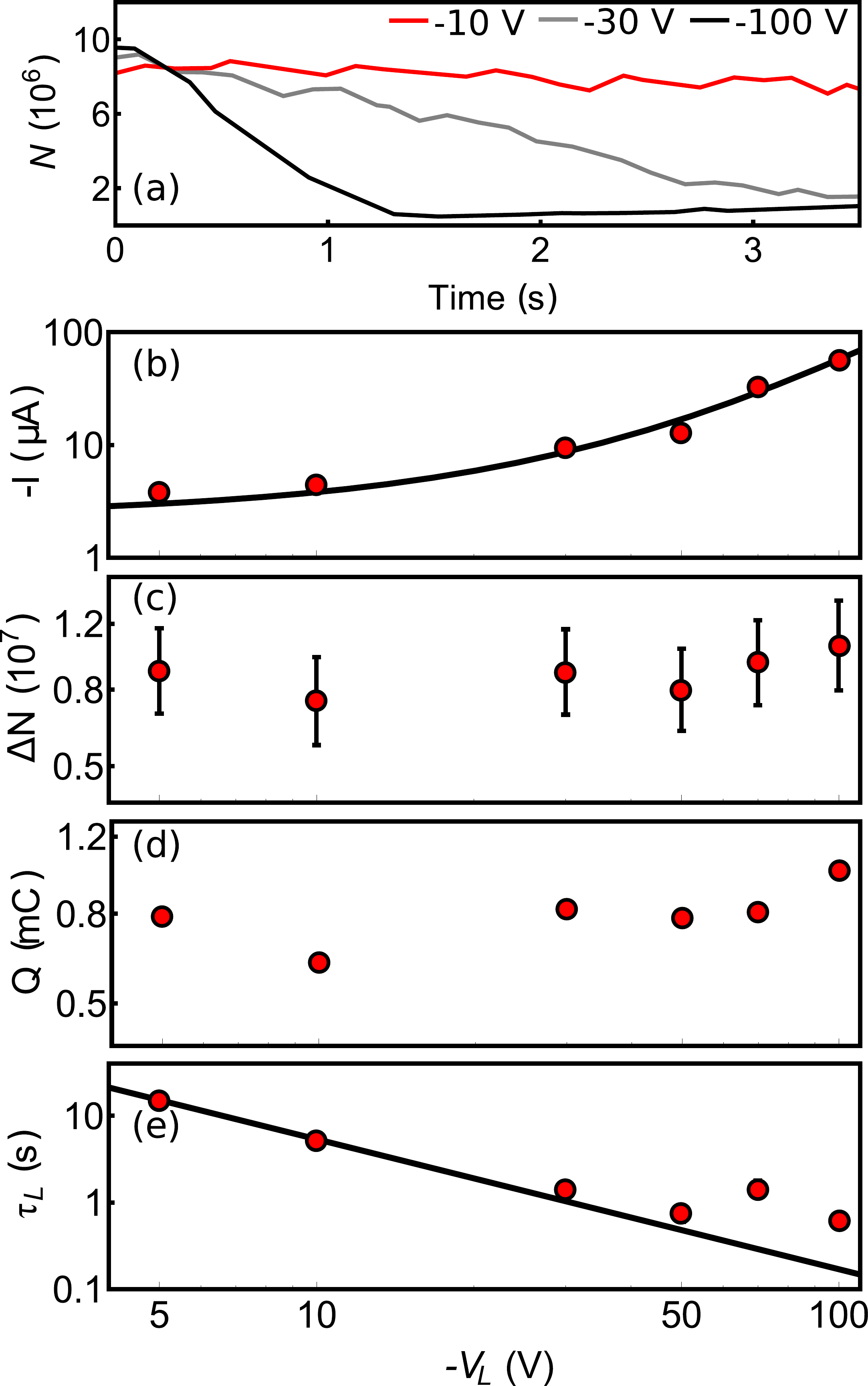}
\caption{The measured parameters for the steady state AIB loading current. (a): The raw time domain data of the atom number during loading with -10~V, -30~V, and -100~V in red, gray, and black respectively. (b): The steady-state current. (c): The change in MOT number during loading. (d): The charge transfer during the first 10~s of loading. (e): The loading time constant, $\tau_{L}$. Black lines indicate the best fit to the data with the functional forms described in the text.}
\label{sinkv}
\end{figure}
If the voltage across the AIB is reversed while trapped atoms are present, the atom number in the MOT decays with the time constant $\tau_{L}$, as shown in Fig.~\ref{sinkv} (a). We measure this time constant as a function of the applied loading voltage by first sourcing the AIB at 30~V until the atom number reached steady-state. The polarity of the applied voltage was reversed, permitting measurement of $\tau_{L}$ and the current drawn by the device as the AIB pumped away the alkali vapor in the chamber. This procedure was carried out for a number of loading voltages, each time ensuring that the starting background density and pressure were constant, as well as the sourcing voltage and steady-state MOT number. Three examples of raw results of this data set are plotted in Fig.~\ref{sinkv} (a), with the full data sets of $\tau_{L}$ as a function of voltage shown in Fig.~\ref{sinkv} (b), (c), (d) and (e). As would be expected from the reaction rate contribution during sourcing, the steady-state loading current is also exponential with the applied voltage, Fig.~\ref{sinkv} (b), indicating both loading and sourcing have a contribution from an electrochemical reaction rate. While the change in atom number $\Delta N$ is nominally independent of the loading voltage due to the similar initial conditions of each experiment, small variation occurred from shot to shot due to slowly varying experimental parameters as a result of environmental fluctuations in the laboratory. We find that the AIB charge transferred during the first 10~s of loading produces a 47~$\%$ fluctuation from the mean, relating to a 34~$\%$ fluctuation with the measured change in cold-atom number, as shown in Fig.~\ref{sinkv} (c) and (d). A correlation coefficient of 0.82, along with the relative insensitivity of the charge transfer to the applied loading voltage, indicates that the number of Rb$^+$ ions that sink in and out of the Rb-$\beta$"-alumina in a given time are proportional to the MOT and background Rb density level present during loading. 
As the loading voltage is increased, a faster electrochemical reaction process is demonstrated for the AIB, consistent with a $\tau_{L}\propto V_L^{-1.5}$ as shown in Fig.~\ref{sinkv} (e), where the black line is a best power law fit. For the largest loading voltages, $\tau_{L}$ appears to saturate with a minimum of $\tau_{L}$~=~540~ms, while the applied voltage continues to increase. This indicates that the process is limited by a time constant prior to the electrochemical disassociation, such as the diffusion of metal Rb upon the AIB upper surface. At the highest loading voltages of -100~V, we were able to sink Rb over 200 times faster than $\tau_{D}$~=~130~s in Fig.~\ref{exampletau} (a), taken with zero applied loading voltage. The short time constant for decreasing the background alkali density could enable rapid loading of a MOT at high background alkali density, followed by long trap lifetimes after subsequent rapid removal of the vapor by the AIB.

\section{Conclusion}

Alkali ion batteries can be used to store, source and sink alkali atoms in instruments based on laser-cooled atomic ensembles. We have demonstrated here that the density of an alkali vapor, and hence the number of atoms in a magneto-optical trap can be controlled by the conditions over which the battery is operated. Increasing the time over which alkali atoms are initially loaded into the AIB results in larger MOT numbers upon subsequent sourcing to the vapor, but also implies longer sourcing time constants. Increasing the sourcing voltage also leads to a larger number of trapped atoms, but with shorter time constants. By combining these parameters it is likely that a long loading time and subsequent large source voltage could provide the best of both worlds, with a large atom number and simultaneous fast time constant. Finally, depletion of atoms from the vapor and MOT depends strongly on the loading voltage and can be made as low as 540~ms under optimal conditions with the current device design. The reversible functionality demonstrated by the device makes an ideal candidate for stabilizing the MOT number in a fluctuating background density or temperature environment \cite{songbai4}. In addition, the AIB is able to preserve the alkali atoms stored in it upon exposure to air and is able to successfully source atoms upon subsequent introduction into a suitable vacuum. 

The results also allow us to elucidate some aspects of the device operation. It appears that both electrochemical reaction and diffusion within the device or on the surface both contribute to the transfer of atoms between the vapor and reservoir. Changes in electrode design may therefore improve the overall device time constants by enabling faster Rb transport on the device surface. It is not yet clear what the capacity of the device to store Rb is since only slight indications of saturation in the MOT atom number were evident for the range of device parameters we explored. The ultimate capacity of the device to store Rb will also be explored in future research.

\section{Acknowledgements}
This material is based upon work supported by the Defense Advanced Research Projects Agency (DARPA) and Space and Naval Warfare Systems Center Pacific (SSC Pacific) (Contract No. N66001-15-C-4027). J.P.M gratefully acknowledges support from the English Speaking Union and Lindemann Fellowship. The authors acknowledge DARPA program manager Robert Lutwak as well as Logan Sorenson, Matthew Rakher, Jason Graetz, John Vajo, Adam Gross, and Danny Kim of HRL Laboratories, LLC for useful discussions. We further acknowledge Florian Herrault, Geovanni Candia, Stephen Lam, Tracy Boden, Margie Cline, Ryan Freeman, and Lian-Xin Coco Huang for assistance with device fabrication. We thank Vincent Maurice and Yun-Jhih Chen for their comments on the manuscript.

\section{Author Contributions}
JPM, KRM and SK conceived and carried out the experiments, supervised by JK and EAD. JPM wrote the manuscript with critical input from KRM, JK, EAD, and CR. CR conceived and designed the alkali-ion battery. RM, AM, and CR fabricated the alkali-ion battery. All authors reviewed the manuscript.

\section{Additional Information}
The authors declare they have no competing interests. Approved for Public Release, Distribution Unlimited.

\end{document}